\documentclass[preprint2,letter]{aastex}

\usepackage{graphicx}
\usepackage[usenames]{color}

\shorttitle{VIMOS Spectroscopy of an E+A Galaxy}
\shortauthors{Chilingarian, De Rijcke, Buyle}

\begin{document}

\title{Internal Kinematics and Stellar Populations of the
  Poststarburst+AGN Galaxy SDSS J230743.41+152558.4\footnote{Based
    on observations made with ESO Telescopes at the La Silla or
    Paranal Observatories under programme ID 077.B-0657}}

\author{I. Chilingarian\altaffilmark{2}}
\email{Igor.Chilingarian@obspm.fr} \author{S. De
  Rijcke\altaffilmark{3}} \email{Sven.DeRijcke@UGent.be}

\and

\author{P. Buyle\altaffilmark{3}}
\email{Pieter.Buyle@UGent.be}

\altaffiltext{2}{Observatoire de Paris-Meudon, LERMA, UMR 8112, 61 Av. de l'Observatoire, 75014 Paris, France; Sternberg Astronomical Institute, Moscow State University, 13 Universitetski prospect, 119992 Moscow, Russia}
\altaffiltext{3}{Astronomical Observatory, Ghent University, Krijgslaan 281, S9, B-9000 Gent, Belgium}

\begin{abstract} 
We present the first 3D spectroscopic observations of a nearby H{\sc
  i} detected poststarburst, or E+A, galaxy, \objectname{SDSS
  J230743.41+152558.4}, obtained with the VIMOS IFU spectrograph at
ESO VLT. Using the {\sc NBursts} full spectral fitting technique, we
derive maps of stellar kinematics, age, and metallicity out to 2--3
half-light radii. Our analysis reveals a large-scale rapidly rotating
disc ($v_{\rm circ} = 300$~km~s$^{-1}$) with a positive age gradient
(0.6 to 1.5~Gyr), and a very metal-rich central region
([Fe/H]=+0.25~dex). If a merger or interaction is responsible for
triggering the starburst, the presence of this undisturbed disc
suggests a minor merger with a gas-rich satellite as the most
plausible option, rather than a disruptive major merger. We find
spectroscopic evidence for the presence of a LINER or AGN. This is an
important clue to the feedback mechanism that truncated the
starburst. The presently observed quiescent phase may well be a
temporary episode in the galaxy's life. \objectname{SDSS
  J230743.41+152558.4} is gas-rich and may restart forming stars,
again becoming blue before finally settling at the red sequence.
\end{abstract}


\section{Introduction}

In the nearby universe, galaxies have a bimodal color distribution,
with a blue peak of star forming galaxies separated from a red
sequence of quiescent galaxies by a sparsely populated gap
\citep{Strateva+01,BGB04,BBN04}. The number of red galaxies has
roughly doubled since $z \sim 1$ \citep{Bell+04}. Red galaxies, on
average, are more luminous than blue galaxies. Therefore, a simple
cessation of star formation in a fraction of the blue galaxies
followed by fading and reddening cannot explain this build-up of the
red sequence. Starbursts are often suggested as a way of increasing a
blue galaxy's stellar mass before letting it fade onto the red
sequence \citep{Bell+04,lfr07}, with mergers and interactions as
possible starburst triggers \citep{bekki05,dimatteo08}. Mergers and
starbursts have the additional appeal that they offer an explanation
for the cessation of star formation : gas is being consumed by the
starburst \citep{dMCMS07} and is being expelled by AGN or supernova
feedback \citep{buyle08,KKSS07}. Thus, previously blue galaxies turn
red and, through further dry mergers, can lead to the formation of the
most massive elliptical galaxies observed in the nearby universe
\citep{Bell+06,nkb06}. This prompted us to investigate the crucial but
transient poststarburst phase, during which a galaxy crosses the color
gap, of this particular evolutionary pathway.

Poststarburst galaxies (PSGs, or E+A galaxies) have optical spectra
with strong Balmer absorption lines, revealing young stars, but faint
if any emission lines, ruling out major star formation
\citep{DG87,CS87,DSP99}. PSGs offer us a unique window on how
starbursts affect galaxy evolution. The starburst population is fading
but still quite prominent and we can study both the burst and the
underlying intermediate-age and/or old stellar populations. Due to
their high surface brightness, we can study their stellar kinematics
out to several half-light radii ($r_e$). Clearly, the poststarburst
phase is of crucial importance to furthering our understanding of the
photometric evolution of the galaxy population.

In this {\it Letter}, we present for the first time two-dimensional
spatially-resolved kinematics and stellar population parameters for a
gas-rich PSG, selected from the \cite{Goto+03} catalog,
\objectname{SDSS J230743.41+152558.4}. With a redshift $z = 0.0695$,
luminosity distance $d_L = 302$~Mpc (adopting a spatially flat
cosmology with $H_0=73$~km~s$^{-1}$~Mpc$^{-1}$, $\Omega_{\rm M}=0.27$,
$\Omega_\Lambda=0.73$), distance modulus $m-M = 37.40$~mag, and a
spatial scale 1.27~kpc~arcsec$^{-1}$, this is a relatively nearby
PSG. The few galaxies with known systemic velocities in a 200~kpc wide
search column centered on \objectname{SDSS J230743.41+152558.4} do not
coincide in redshift and are spread over a very large redshift range
($0.036 \lesssim z \lesssim 0.25$), making it unlikely that they are
physically connected with \objectname{SDSS J230743.41+152558.4}. We
confirm the classification of \objectname{SDSS J230743.41+152558.4} by
\citet{yagi06} as a field PSG.

 This is the first paper in a series based on our VLT/VIMOS
 observations of 3 PSGs.



\section{Observations and Data Reduction} \label{sec:obs}


The data have been obtained with the integral field unit (IFU) mode of
the VIMOS spectrograph at ESO/VLT-Melipal by ESO staff in
mid-2006. With the high-resolution ``HR-blue'' spectral setup, the
0.67$''$ per lens spatial scale provides a field of view (FoV) of
$27'' \times 27''$ (40$\times$40~lenses).  This setup offers a
spectral resolving power $R \sim 2500$ between 4200 and 6300~{\AA}.
In total, we use eleven out of twelve obtained exposures, spatially
dithered in order to work around broken IFU fibers, and each with an
on-source time of 2,000~sec. The seeing FWHM varies between 0.7 and
1.2$''$. 
Arc line and flat field
calibration frames have been taken for every Observing Block (OB).

For the data recuction, we use the generic IFU data reduction pipeline
implemented in {\sc ITT IDL}.  A brief description of the data
reduction steps can be found in \cite{CPSA07}. Detailed information
including all VIMOS-specific aspects concerning these data will be
provided in a forthcoming paper. Each of the four VIMOS quadrants,
labeled Q1 through Q4, of every OB was reduced independent from the
others until the flux calibration step. The relative flux calibration
of the quadrants is performed using twilight sky frames and requires
extra care since we use spatial dithering. 
The absolute flux calibration was done using observations of
spectrophotometric standards included in the standard VIMOS
calibration plan.

Night sky spectra are reconstructed from the lenses outside 5~r$_e$ of
the galaxy to minimize  contamination by the galactic stellar
continuum and are subtracted from the science frames. Then, individual
exposures are co-added, applying iterative sigma-clipping to remove
cosmic ray hits, spatial and spectral shifts to account for spatial
dithering, different heliocentric corrections and atmosphere
differential refraction. Noise frames are computed using photon
statistics and are processed through the same reduction steps as the
science frames, up to the sky subtraction.

We use twilight and science frames contaminated by Moon (4 out of 12) to
assess the data reduction quality and obtain quantitative estimates of the
VIMOS spectral line spread function variations across the FoV, along the
wavelength axis, and between different OBs. We fit the solar spectrum, taken
from the ELODIE.3.1 stellar library \citep{PSKLB07}, against the observed
spectra in every lens at six wavelength intervals between 4000 and 6300{\AA}
using the {\sc PPXF} procedure \citep{CE04}. From this test, we conclude
that (1) wavelength calibration quality is better than 5~km~s$^{-1}$ between
4000 and 5900\AA; (2) spectral resolution ($\sigma_{\mbox{inst}}$) exhibits
no significant variations across the FoV for individual quadrants, neither
between observing blocks, although (3) it does change along wavelength (60
to 40~km~s$^{-1}$ from blue to red) and between quadrants; (4) high-order
moments of the Gauss-Hermite parametrisation ($h_3$ and $h_4$,
\citealp{vdMF93}) are close to zero except Q3 at $\lambda < 4350$~\AA, where
$h_3$ is modestly negative due to imperfect focusing; (5) systematic errors
of the wavelength calibration become very important at $\lambda > 5900$\AA,
changing significantly between individual OBs so we restrict our analysis to
shorter wavelengths.

\section{Data Analysis} \label{sec:ana}

We use the {\sc NBursts} full spectral fitting technique \citep{CPSK07}
with high-resolution ($R=10,000$) {\sc pegase.hr} \citep{LeBorgne+04} simple
stellar population (SSP) models to extract kinematics and stellar
population from the absorption-line spectra. First, we rebin the data
to a target signal-to-noise ratio of 15 per bin using the Voronoi
adaptive tessellation technique \citep{CC03}. We use only that part of
the VIMOS IFU FoV where the sky-subtraction uncertainty is
sufficiently small, with a size of 18$\times$18 lenses ($12'' \times
12''$). The subsequent fitting procedure comprises the following
steps: (1) a grid of SSP spectra with a fixed set of ages (spaced
nearly logarithmically from 20~Myr to 18~Gyr) and metallicities (from
$-$2.0 to $+$0.5~dex) is convolved with the wavelength-dependent
instrumental response of VIMOS in every Voronoi tessella as explained
in \cite{CPSA07}; (2) a non-linear least square fitting against an
observed spectrum is done for a template picked from the pre-convolved
SSP grid using 2D-spline interpolation on age ($\log t$), and
metallicity ($Z$), broadened with the line-of-sight velocity
distribution (LOSVD) parametrized by $v$, $\sigma$, $h_3$, and $h_4$
and multiplied pixel-by-pixel by an $n^{\rm{th}}$ order Legendre
polynomial (multiplicative continuum), resulting in $n + 7$ parameters
to be determined (we use $n=25$). 

We exclude narrow 12\AA-wide regions around the H$\beta$, H$\gamma$,
[O{\sc iii}] ($\lambda=4959, 5007$\AA), and [N{\sc i}]
($\lambda=5199$\AA) lines from the fit in order to be able to detect faint emission
lines. As shown in \citet{CPSA07,Chilingarian+08}, reliable
SSP-equivalent age and metallicity estimates are obtained even if the
Balmer absorption lines are excluded. Using the approach described in
\citet{Chilingarian09} we estimate the excluded regions to contain
only $\sim 7$\% of age-sensitive information.


\section{Results} \label{sec:res}

\begin{figure}
\begin{tabular}{ccc}
\includegraphics[width=0.5\hsize]{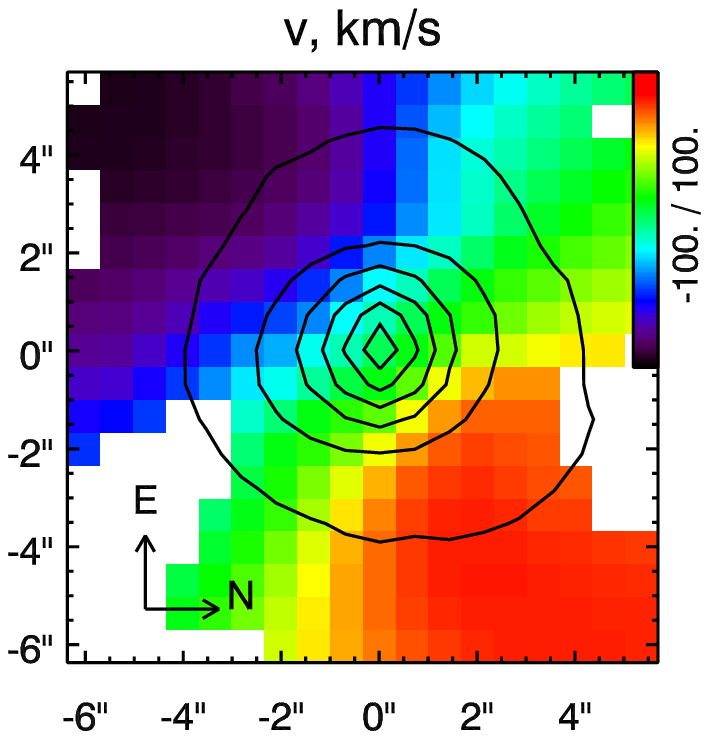} &
\includegraphics[width=0.5\hsize]{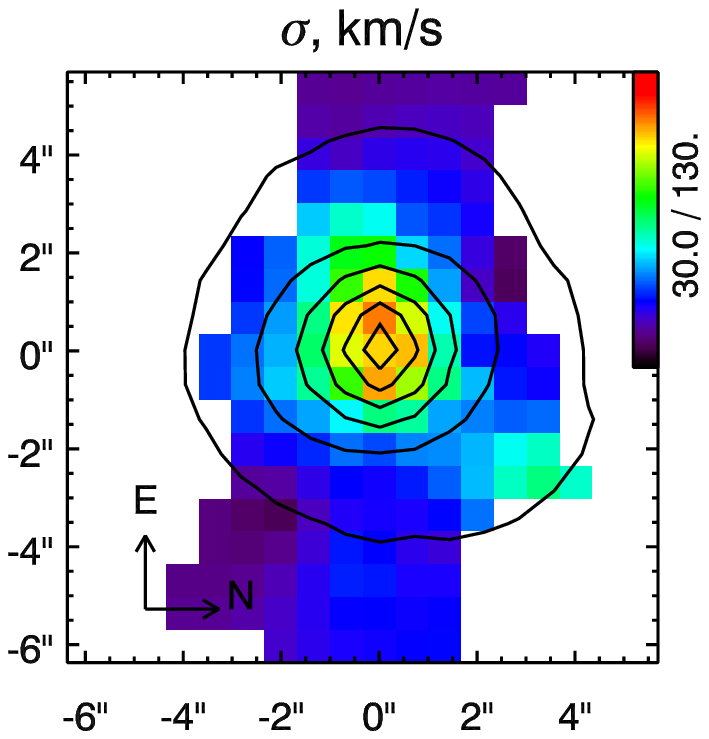} \\
\includegraphics[width=0.5\hsize]{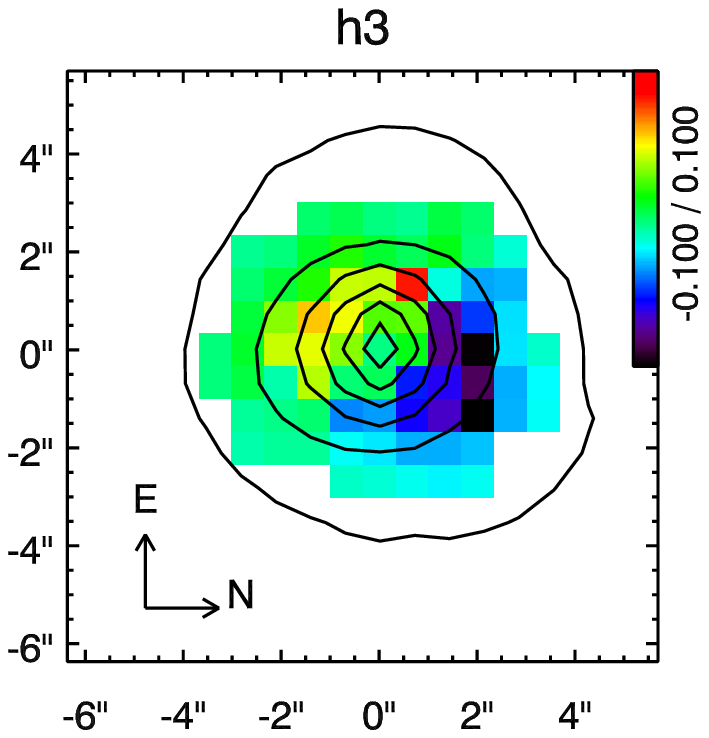} &
\includegraphics[width=0.5\hsize]{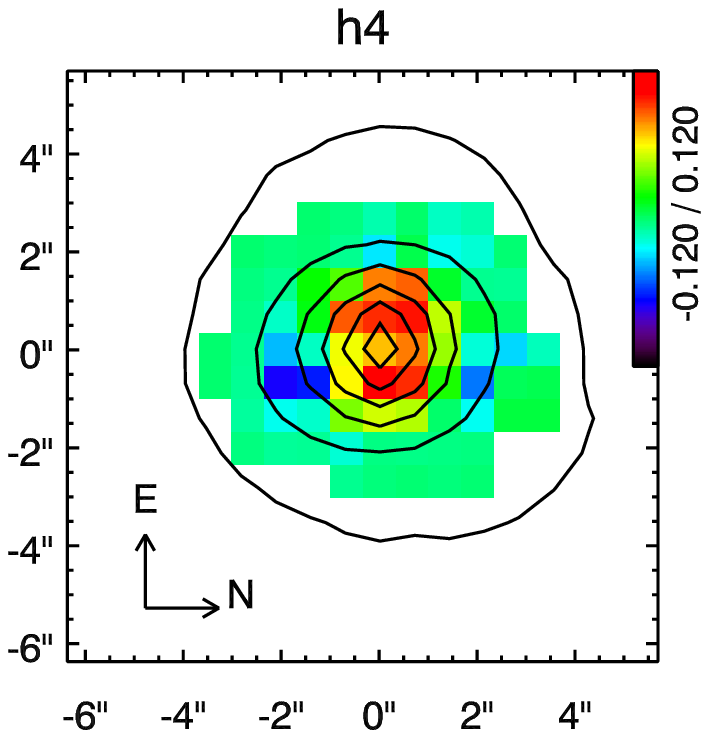} \\
\includegraphics[width=0.5\hsize]{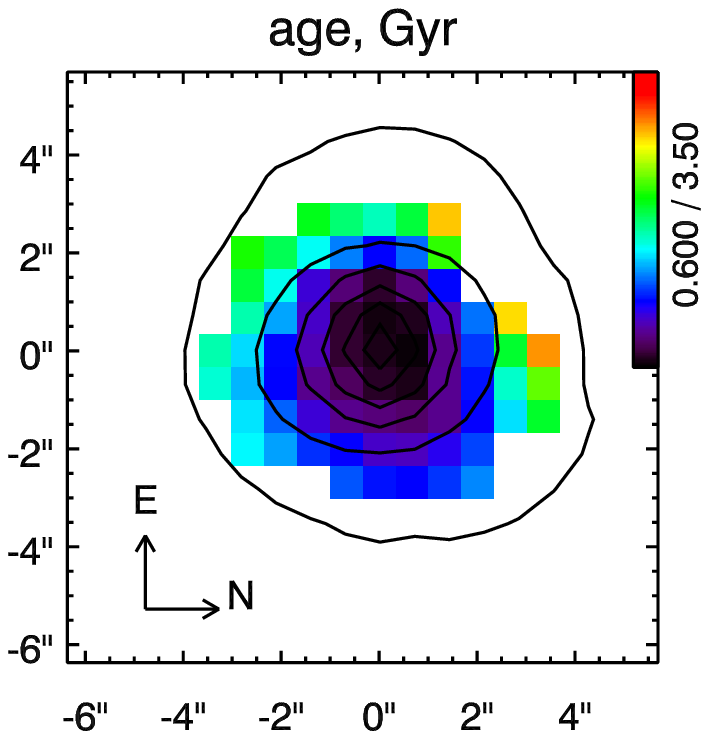} &
\includegraphics[width=0.5\hsize]{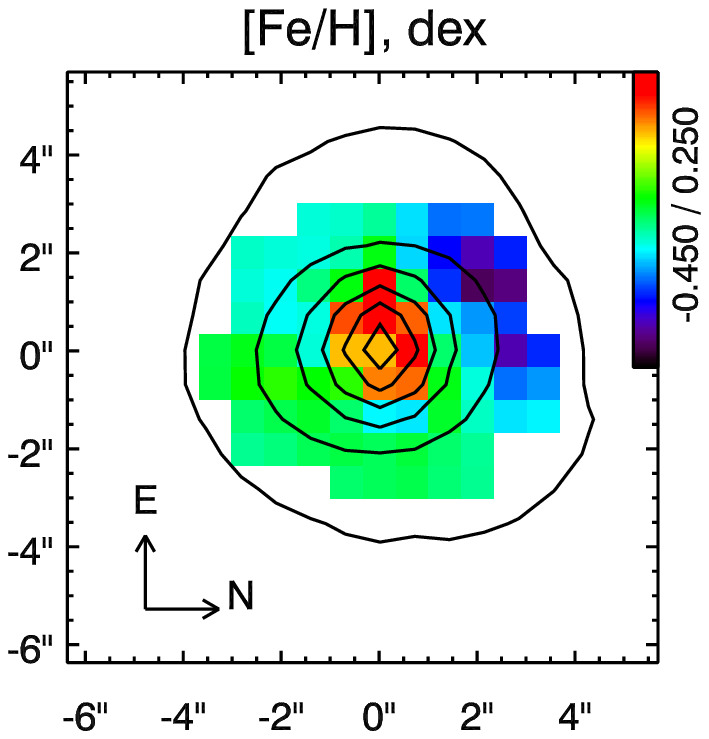} \\
\end{tabular}
\caption{Two-dimensional maps of stellar kinematics, SSP-equivalent
  age and metallicity for SDSS~J230743.41+152558.4 derived from the
  VIMOS IFU data. The black contours indicate isophotes of the
  reconstructed image, obtained by collapsing the data cube along the
wavelength axis. \label{figmaps}}
\end{figure}

We present the maps derived from the full-spectrum fitting in
Fig~\ref{figmaps}. Those regions where the parameters have large
uncertainties (e.g. $>15$~km~s$^{-1}$ for $v$ and $\sigma$) or are
not fit ($h_3$ and $h_4$ in the outer regions) are masked. The
isophotes from the reconstructed continuum image (datacube collapsed
along the spectral dimension) are shown as black solid
contours. Line-of-sight velocities can be determined with adequate
precision out to larger radii from the galaxy centre than the velocity
dispersion and the stellar population parameters, because they are
insensitive to the sky subtraction.

\subsection{Kinematics and dynamics}

\begin{figure}
\includegraphics[width=\hsize]{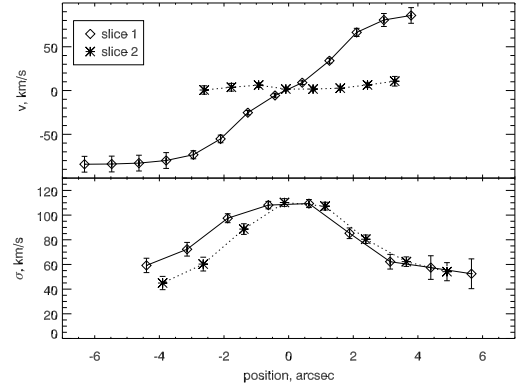}
\caption{The line-of-sight velocity $v$ and velocity dispersion
  $\sigma$ along two cuts through SDSS~J230743.41+152558.4, one
  tracing the kinematical minor axis, the other the kinematical major
  axis. The velocity profile flattens out around $\sim 5''$ while
  the velocity dispersion drops to $\sigma \sim 50$~km~s$^{-1}$.\label{figrvpr}}
\end{figure}

\objectname{SDSS J230743.41+152558.4} exhibits very regular, symmetric
pure disc rotation with a semi-amplitude of 90~km~s$^{-1}$ without any
evidence for significant isovelae twist. However, there is a
significant misalignment of the kinematical and photometrical axes. We
mention the evident large-scale bar oriented exactly along the
galaxy's minor axis (roughly North-South), which is visible on SDSS
images and also noticeable here in the reconstructed image. 
The rotation curve flattens out at $\sim 4''$ from the centre,
corresponding to $\sim 6$~kpc or $\sim 3.5 r_e$ \citep{Buyle+06} (see
Fig. \ref{figrvpr}). The velocity dispersion declines from a central
maximum of $\sim$110~km~s$^{-1}$ to $40-50$~km~s$^{-1}$ at 6$''$. The
rather high $v/\sigma$ in the outskirts of the galaxy also argue for
the presence of a disc. One has to be careful with interpreting the
high very central values, because a rapidly rotating circumnuclear
component smeared by atmospheric seeing may produce very similar
effects. Weighing all the evidence, \objectname{SDSS
  J230743.41+152558.4} is best classified morphologically as a barred
lenticular or SB0 galaxy. This is not surprising since, although the
diversity is quite large, many E+As have S0-like morphologies
\citep{y08}.

We use a technique similar to that proposed by \cite{vMW85} to
determine the galaxy's inclination, $i$, and the position angle of the
major axis, PA, from the stellar radial velocity field (the main
difference being the non-parametric representation of the rotation
curve). Fitting the velocity field in elliptical annuli in the radial
range $0.7< r < 6''$ gives PA$=131$~deg and $i=30$~deg. These values
agree well with the shape and axial ratio of the outer isophote
displayed in Fig~\ref{figmaps}. The peak projected velocity $v \approx
90$~km~s$^{-1} = v_\phi \sin i$, after correcting for inclination
effects, yields $v_\phi \approx 180$~km~s$^{-1}$ for the mean
tangential velocity. We adopt the value $h_R=0.8''$, provided by SDSS,
for the exponential scale-length of the stellar density profile. Using
these numbers, we estimate the circular velocity $v_{\rm circ}$ at a
radius of 4$''$, corrected for asymmetric drift using the method
outlined in paragraph 4.8.2(a) of \citet{bt08}.  For the outskirts of
the galaxy, we assume the rotation curve to be flat ($\sigma_\phi
\approx \sigma_R/\sqrt{2}$) and the velocity dispersions to be
independent of radius. We then arrive at the following expression for
the circular velocity:
\begin{equation}
v_{\rm circ}(4'') \approx \sqrt{ v_\phi^2 + \sigma_\phi^2\left(2
  \frac{4''}{h_R}-1\right)}.
\end{equation}
Here, $\sigma_\phi$ is the tangential velocity
dispersion. Unfortunately, along the major axis, $\sigma$ is a
function of both $\sigma_\phi$ and $\sigma_z$, the vertical component
of the velocity dispersion tensor. If we assume we are dealing with a
thin disc then $\sigma_z \ll \sigma_\phi$ and $\sigma_\phi \approx
\sigma/\sin i \sim 80$~km~s$^{-1}$. In the case of an isotropic
velocity dispersion tensor, $\sigma_\phi = \sigma_z \approx \sigma
\sim 40$~km~s$^{-1}$. The first case yields $v_{\rm circ} \sim
300\,{\rm km}\,{\rm s}^{-1}$; the latter leads to $v_{\rm circ} \sim
220\,{\rm km}\,{\rm s}^{-1}$. The velocity width (W20) of the H{\sc i}
21~cm emission of this galaxy is $286 \pm 43$~km~s$^{-1}$. This
corresponds to a circular velocity $v_{\rm circ} \approx 143/ \sin
i$~km~s$^{-1} \sim 286$~km~s$^{-1}$ (Buyle et al., in prep.), sitting
comfortably between the two extreme values derived from stellar
kinematics. We will adopt it here as our best-guess value. Using these
numbers, we estimate the total dynamical mass within a 5~kpc radius at
$\sim 10^{11}$~M$_\odot$.

With a total B-band absolute magnitude M$_B = -20.5$~mag (Buyle et
al., in prep.), logarithm of the peak circular velocity $\log(v_{\rm
  circ})= 2.48$, logarithm of the central velocity dispersion
$\log(\sigma_0)=2.04$, \objectname{SDSS J230743.41+152558.4} sits
exactly on the Tully-Fisher (\citeyear{TF77}) relation for elliptical
and S0 galaxies \citep{bam06,sdr07}. It is slightly offset from the
E/S0 Faber-Jackson relation \citep{FJ76,ortkro01,deRijcke+05}, being
almost $\sim 1$~mag too bright in the B-band for its velocity
dispersion.

In Fig.~\ref{figFP}, we present the Fundamental Plane (FP,
\citealp{DD87}) in $\kappa$-space \citep{BBF92}. 
On the $\kappa_2$~vs.~$\kappa_1$ plot (the plane's ``face-on view''),
the distinct regions are occupied by dwarf
\citep{vZSH04,GGvdM03,deRijcke+05,Chilingarian+08} and intermediate
luminosity and giant early-type galaxies \citep{BBF92}. The latter
region is extended towards the upper-left corner of the plot by
low-luminosity bulges and compact elliptical galaxies with M59cO
\citep{CM08} being the most extreme case. \objectname{SDSS
  J230743.41+152558.4}, indicated by the red filled circle, falls in
the region occupied by high-surface brightness low-luminosity bulges
and ellipticals. Since the FP is defined for early-type galaxies
(E/S0) assumed to be virialised systems, we should include the
rotational energy into the total balance at a first approximation as
$\sigma^2 + v_{r}^2/2$. This moves \objectname{SDSS
  J230743.41+152558.4} to the position indicated by the end of the
green vector in Fig~\ref{figFP}, toward the locus of bright E/S0
galaxies.

\subsection{Stellar populations and nuclear activity}

In the stellar population maps, we clearly see a barely resolved core
with a shape and size similar to that of the central
$\sigma$-bump. The galaxy center reaches a significantly super-solar
metallicity, up to $+0.25$~dex. The central SSP-equivalent age is
650~Myr, increasing to approximately 3~Gyr at 3$''$ from the galaxy
nucleus. The starburst region, where the very young population
resides, is about 2~arcsec (2.5~kpc) across. \citet{yagi06} observed
\objectname{SDSS J230743.41+152558.4} using long-slit spectroscopy and
found large H$\delta$ equivalent widths out to 2.5~arcsec. The VIMOS
data confirm the existence of an extended region showing a young
stellar population, although the positive age gradient is
significant. For a comparison with the SDSS DR6 spectrum
\citep{sdssdr6} of this galaxy, we integrate the light in the
data-cube inside a 3$''$ circular aperture and apply the full-spectrum
fitting technique to both spectra. We find excellent agreement:~all
the kinematical and stellar population parameters, including $h_4$,
are consistent within the uncertainties.

\begin{figure*}
\begin{tabular}{cc}
\includegraphics[width=0.2\hsize]{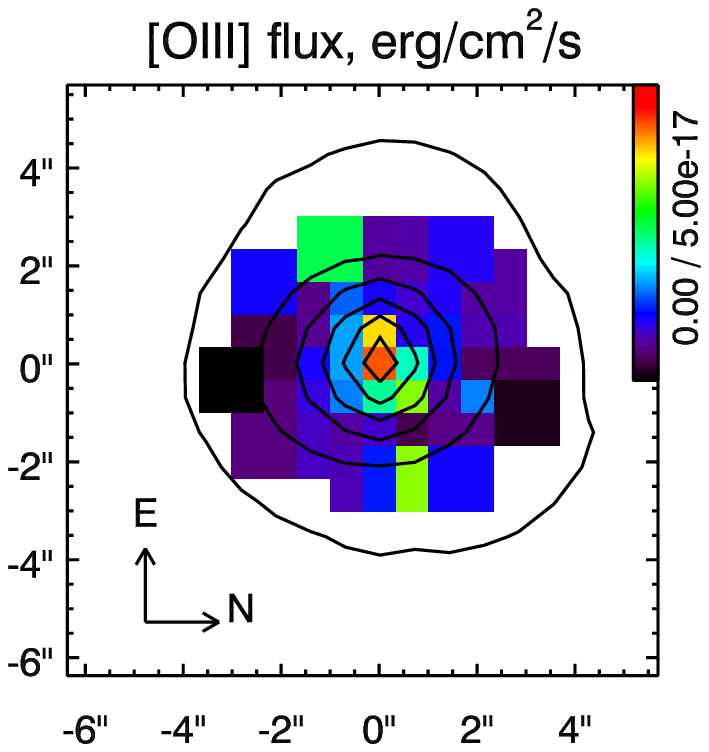} &
\includegraphics[width=0.8\hsize]{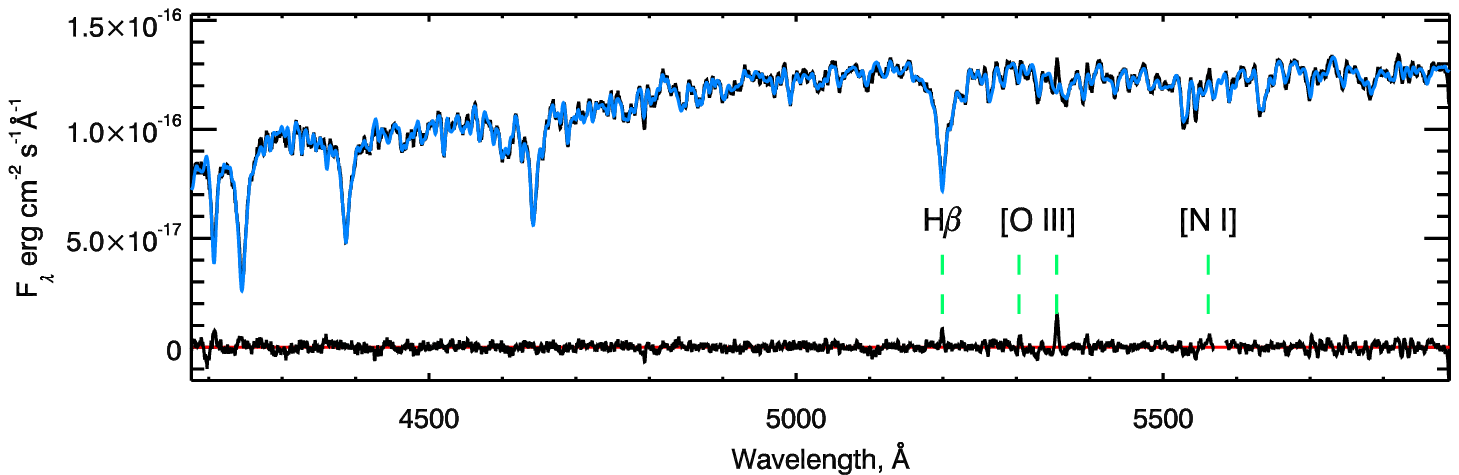} \\
\end{tabular}
\caption{Distribution of [OIII] emission in the central region of SDSS J230743.41+152558.4
(left) and the integrated spectrum of the inner 2$''$ (right).
\label{figspec}}
\end{figure*}

We fit the brightest emission line ([O{\sc iii}] $\lambda=5007$\AA)
with a Gaussian to determine its amplitude (see
Fig~\ref{figspec}a). Unfortunately, no precise measurements of radial
velocities can be made due to the very low emission-line flux. The
[O{\sc iii}] emission is very centrally concentrated, arguing for the
presence of a spatially unresolved emitting region. Careful inspection
of the fitting residuals in the central region (Fig~\ref{figspec}b)
reveals barely-detectable emission in the forbidden [N{\sc i}] line
($\lambda = 5199$~\AA), which is a typical signature of a LINER or
AGN. We therefore connect the central peak of the [O{\sc iii}]
distribution to possible nuclear activity. Fitting the SDSS DR6
spectrum in the full wavelength range of {\sc pegase.hr} ($3900 <
\lambda < 6800$~\AA) and subtracting the best-fitting template from
the data reveals strong emission lines in H$\alpha$ and [N{\sc ii}]
($\lambda = 6548, 6584$\AA). The high $\log ([$N{\sc
    ii}$_{6584}$]/H$\alpha)=0.13$ and $\log ([$O{\sc
    iii}$_{5007}$]/H$\beta)=0.46$ ratios support the LINER
interpretation \citep{BPT81,kewley06}. This would make
\objectname{SDSS J230743.41+152558.4} the second well-studied PSG to
date in which low-power nuclear activity has been detected
\citep{liu07}. As proposed by \citet{KKSS07}, AGN feedback may play a
crucial role in quenching star formation in massive PSGs by expelling
the gas. \citet{buyle08} find that most of the H{\sc i} gas in the
binary PSG system EA01A/B resides {\em outside} the stellar bodies of
the galaxies, suggesting a feedback process powerful enough to
physically displace large quantities of gas, such as AGN feedback
\citep{silkrees98,schawinski09}. At the same time, the faint H$\beta$
and [O{\sc iii}] ($\lambda=4959${\AA}) emission lines are detected
almost everywhere out to $3''$ from the galaxy centre, suggesting weak
ongoing star formation over a large part of the galaxy.

\section{Past and future evolution} \label{sec:dis}

\begin{figure}
\includegraphics[width=\hsize]{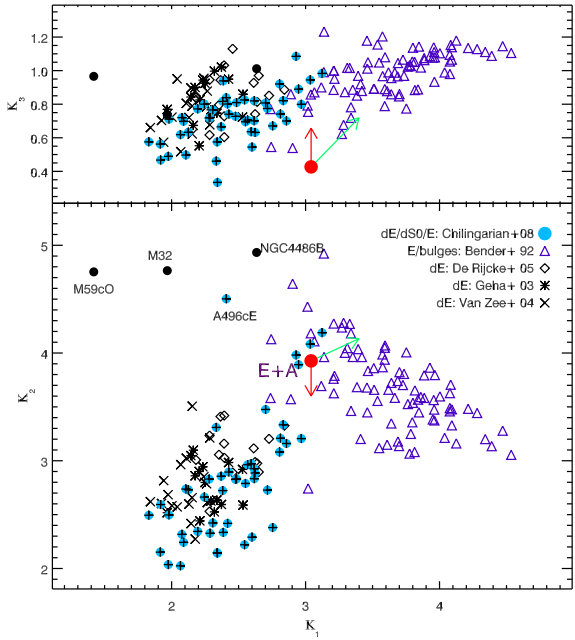}
\caption{$\kappa$-space view of the Fundamental Plane. The
position of SDSS J230743.41+152558.4 is shown with a filled red circle. See text for other sources of
data. The end of the red vector indicates the position of SDSS J230743.41+152558.4 after 900~Myr
of passive evolution while the green vector denotes its shift when taking
the rotational energy into account.\label{figFP}}
\end{figure}

The distance to \objectname{SDSS J230743.41+152558.4} corresponds to a
light travel time of $\sim900$~Myr. We estimate its $B$-band
luminosity evolution using the PEGASE.2 \citep{FR97} evolutionary
synthesis code. If we imagine \objectname{SDSS J230743.41+152558.4}
evolving passively during this period, it will fade by, at maximum,
1~mag if all its stars formed 700~Myr ago. In the more plausible case
of a composite stellar population, the effect on the total $B$
magnitude will be even smaller, because the old population evolves
slower than the young one, which has a lower mass fraction. This will
make \objectname{SDSS J230743.41+152558.4} resemble present-day
intermediate-luminosity early-type galaxies (E/S0) in its position on
the Faber-Jackson relation, the Kormendy relation \citep{Kormendy77},
and the Fundamental Plane (see red vector in Fig~\ref{figFP}). 4$''$,
or $3 r_e$, encloses about 75~\% of the light, resulting in a
mass-to-light ratio of $M/L_B \sim 4$ in solar units. After 700~Myr of
passive evolution, this will have increased to $M/L_B \sim 10$, at
maximum. This is compatible with present-day early-type galaxies
\citep{ortkro01}.

\objectname{SDSS J230743.41+152558.4} turns to be a fast rotator if one
applies the recently proposed classification \citep{Emsellem+07}. However,
despite its regular morphology, given the presence of a dominating    
large-scale dynamically cold disc, we should not classify this object as a
typical early-type galaxy. Major mergers and interactions may lead to
strong, abruptly truncated star formation episodes, as shown statistically
using numerical simulations by \cite{dMCMS07}, resulting in the E+A   
phenomenon. Such violent events heat and often even completely         
destroy the discs, leaving little opportunity to explain the observed  
velocity field of \objectname{SDSS J230743.41+152558.4}. On the other hand,
a minor merger of a large disc-dominated lenticular or early-type spiral
with an intermediate-mass gas-rich satellite may be an acceptable
explanation: young stars will actively form in the disc from the accreted
intermediate-metallicity ISM while the central star formation peak will
consume the metal-rich gas often present in the circumnuclear regions of
early-type galaxies.

Given the presence of neutral gas \citep{Buyle+06}, the quenching of
star formation in \objectname{SDSS J230743.41+152558.4} could very
well be a transient phase. For instance, the gaseous component could
have been dispersed by supernova or AGN feedback and may re-accrete
after several hundred Myrs, possible restarting the star formation. If
this object had been observed during a star-forming episode, it would
have been classified as a barred late-type spiral (SBcd).


Deep H{\sc i} radio observations have led to the detection of 21~cm
emission in about 60~per~cent of the targeted PSGs
\citep{Buyle+06}. Some of the detected PSGs are as gas-rich as
local LIRGS and mergers, possible progenitors of PSGs in the local
universe \citep{KKSS07}. The presence of large quantities of gas
indicates that many PSGs might actually be observed during an inactive
phase of the star-formation duty cycle. This suggests that the
evolution from blue to red galaxy may encompass several cycles back
and forth in the colour--magnitude diagram before the galaxy finally
settles on the red sequence.

\acknowledgments

IC acknowledges the RFBR grant 07-02-00229-a.

Our study makes use of SDSS DR6. Funding for the Sloan Digital Sky
Survey (SDSS) and SDSS-II has been provided by the Alfred P. Sloan
Foundation, the Participating Institutions, the National Science
Foundation, the U.S. Department of Energy, the National Aeronautics
and Space Administration, the Japanese Monbukagakusho, and the Max
Planck Society, and the Higher Education Funding Council for
England. The SDSS Web site is http://www.sdss.org/.

The SDSS is managed by the Astrophysical Research Consortium (ARC) for
the Participating Institutions. The Participating Institutions are the
American Museum of Natural History, Astrophysical Institute Potsdam,
University of Basel, University of Cambridge, Case Western Reserve
University, The University of Chicago, Drexel University, Fermilab,
the Institute for Advanced Study, the Japan Participation Group, The
Johns Hopkins University, the Joint Institute for Nuclear
Astrophysics, the Kavli Institute for Particle Astrophysics and
Cosmology, the Korean Scientist Group, the Chinese Academy of Sciences
(LAMOST), Los Alamos National Laboratory, the Max-Planck-Institute for
Astronomy (MPIA), the Max-Planck-Institute for Astrophysics (MPA), New
Mexico State University, Ohio State University, University of
Pittsburgh, University of Portsmouth, Princeton University, the United
States Naval Observatory, and the University of Washington.




{\it Facilities:} \facility{ESO VLT (VIMOS)}.




\clearpage



\end{document}